\documentclass[a4paper,12pt]{article}
\usepackage{comment}
\usepackage{cite}
\usepackage{amsmath}
\usepackage{amssymb}
\usepackage{amsfonts}
\usepackage[T1]{fontenc}
\usepackage[utf8]{inputenc}
\usepackage{graphicx}
\usepackage{fancyhdr}
\usepackage{float}
\usepackage{xcolor}
\usepackage{authblk}
\usepackage{mathrsfs}
\usepackage{empheq}

\usepackage{geometry}
\begin{document}

\title{Liquid-gas transition in nuclear matter:\\
analytical formulas for the Virial coefficients}

\author{Jean-Christophe {\sc Pain}\\
\small
CEA, DAM, DIF, F-91297 Arpajon, France\\
Université Paris-Saclay, CEA, Laboratoire Matière en Conditions Extr\^emes,\\ 
91680 Bruyères-le-Ch\^atel, France
}

\maketitle

\large

\begin{center}

\end{center}

\normalsize

\begin{abstract}
In many fields of statistical physics, for instance in the study of the liquid-gas phase transition in finite nuclear matter, the Virial coefficients of the Fermi gas play a major role. In this note, we provide relations, sum rules, analytical formulas and numerical values for such coefficients\footnote{This work was suggested in 1995-1996 by Jacques Meyer (Professor at University Claude Bernard Lyon 1).}.
\end{abstract}

\vspace{1cm}
\noindent {\bf Keywords: liquid-gas transition; Skyrme interaction; Virial coefficients; Fermi ideal gas; critical point.} 

\clearpage
\setcounter{page}{3}
\clearpage

\section{Introduction}

Using the finite-temperature Hartree-Fock theory, as presented by Fetter and Walecka \cite{Fetter}, it is possible to derive an equation of state for a Fermi gas of nucleons interacting through the Skyrme \cite{Su1987} force. Details of the calculation are provided in an appendix of Ref. \cite{Jaqaman1983}, but the resulting equation is quite simple:
\begin{equation}
P=-a_0\rho^2+a_3(1+\sigma)\rho^{2+\sigma}+\left(1-\frac{3}{2}\frac{\rho}{m^*}\frac{dm^*}{d\rho}\right)P_{\mathrm{id}}(m^*),    
\end{equation}
where $P_{\mathrm{id}}(m^*)$ is the pressure of a Fermi ideal gas made of particles with mass $m^*$ at the temperature $T$. It can be obtained \emph{via} the Virial expansion:
\begin{equation}\label{Pid}
P_{\mathrm{id}}=k_BT\sum_{n=1}^{\infty}B_n\rho^n,
\end{equation}
where $B_n$ are the so-called Virial coefficients. One has
\begin{equation}\label{pres}
\frac{P_{\mathrm{id}}}{k_BT}=\frac{g}{\lambda^3}f_{5/2}(z),
\end{equation}
$g$ being the spin-isospin degeneracy factor and
\begin{equation}
\lambda=\left(\frac{2\pi\hbar^2}{m^*k_BT}\right)^{1/2}
\end{equation}
the thermal de Broglie wavelength. The $f_{5/2}(z)$ Fermi function reads
\begin{equation}
f_{5/2}(z)=\frac{4}{\sqrt{\pi}}\int_0^{\infty}x^2\ln\left(1+ze^{-x^2}\right)dx
\end{equation}
and can be expanded as
\begin{equation}
f_{5/2}(z)=\sum_{n=1}^{\infty}(-1)^{n+1}\frac{z^n}{n^{5/2}}.
\end{equation}
The density of the Fermi ideal gas reads
\begin{equation}
\rho=\frac{g}{\lambda^3}f_{3/2}(z)
\end{equation}
with
\begin{equation}
f_{3/2}(z)=z\frac{\partial}{\partial z}f_{5/2}(z)
\end{equation}
satisfying the expansion
\begin{equation}
f_{3/2}(z)=\sum_{n=1}^{\infty}(-1)^{n+1}\frac{z^n}{n^{3/2}}.
\end{equation}
Combining Egs. (\ref{Pid}) and (\ref{pres}), one gets
\begin{equation}\label{equaares}
\frac{g}{\lambda^3}\sum_{n=1}^{\infty}(-1)^{n+1}\frac{z^n}{n^{5/2}}=\sum_{n=1}^{\infty}B_n\left(\sum_{k=1}^{\infty}(-1)^{k+1}\frac{z^k}{k^{3/2}}\right)^n
\end{equation}

\section{Direct ``brute force'' calculation}

Equation (\ref{equaares}) is equivalent to

\begin{eqnarray}
\frac{g}{\lambda^3}\sum_{n=1}^{\infty}(-1)^{n+1}\frac{z^n}{n^{5/2}}&=&B_1\frac{g}{\lambda^3}\sum_{n=1}^{\infty}(-1)^{n+1}\frac{z^n}{n^{3/2}}\nonumber\\
& &+B_2\frac{g^2}{\lambda^6}\left(\sum_{n=1}^{\infty}(-1)^{n+1}\frac{z^n}{n^{3/2}}\right)^2+\cdots
\end{eqnarray}
and thus
\begin{eqnarray}
\sum_{n=1}^{\infty}(-1)^{n+1}\frac{z^n}{n^{5/2}}&=&B_1\sum_{m=1}^{\infty}(-1)^{m+1}~\frac{z^m}{m^{3/2}}\nonumber\\
& &+B_2\frac{g}{\lambda^3}\sum_{m,n\ge 1}^{\infty}(-1)^{m+n+2}~\frac{z^{m+n}}{(mn)^{3/2}}\nonumber\\
& &+B_3\frac{g^2}{\lambda^6}\sum_{m,n,p\ge 1}^{\infty}(-1)^{m+n+p+3}~\frac{z^{m+n+p}}{(mnp)^{3/2}}\nonumber\\
& &+B_4\frac{g^3}{\lambda^9}\sum_{m,n,p,q\ge 1}^{\infty}(-1)^{m+n+p+q+4}~\frac{z^{m+n+p+q}}{(mnpq)^{3/2}}\nonumber\\
& &+B_5\frac{g^4}{\lambda^{12}}\sum_{m,n,p,q,r\ge 1}^{\infty}(-1)^{m+n+p+q+r+5}~\frac{z^{m+n+p+q+r}}{(mnpqr)^{3/2}}\nonumber\\
& &+B_6\frac{g^5}{\lambda^{15}}\sum_{m,n,p,q\ge 1}^{\infty}(-1)^{m+n+p+q+r+s+6}~\frac{z^{m+n+p+q+r+s}}{(mnpqrs)^{3/2}}\nonumber\\
& &+B_7\frac{g^6}{\lambda^{18}}\sum_{m,n,p,q,r,s,t\ge 1}^{\infty}(-1)^{m+n+p+q+r+s+t+7}~\frac{z^{m+n+p+q+r+s+t}}{(mnpqrst)^{3/2}}\nonumber\\
& &+\cdots.
\end{eqnarray}
Identification of the powers of $z$ yields the Virial $B_n$ coefficients.

\vspace{5mm}
For $n=1$:
\begin{equation}
\frac{1}{1^{5/2}}=B_1\frac{1}{1^{3/2}} \Rightarrow B_1=1.
\end{equation}
For $n=2$: 
\begin{equation}
-\frac{1}{2^{5/2}}=-\frac{B_1}{2^{3/2}}+B_2\frac{g}{\lambda^3}.\frac{1}{1^{3/2}} \Rightarrow B_2=\frac{1}{2^{5/2}}\left(\frac{\lambda^3}{g}\right).
\end{equation}
For $n=3$: 
\begin{equation}
\frac{1}{3^{5/2}}=\frac{B_1}{3^{3/2}}+B_2\frac{g}{\lambda^3}\left(-\frac{2}{2^{3/2}}\right)+B_3\frac{g^2}{\lambda^6}\frac{1}{1^{3/2}}\Rightarrow B_3=\left(\frac{1}{8}-\frac{2}{9\sqrt{3}}\right)\left(\frac{\lambda^3}{g}\right)^2.
\end{equation}
For $n=4$: 
\begin{equation}
B_4=\left(\frac{3\sqrt{6}+5\sqrt{3}-16}{32\sqrt{6}}\right)\left(\frac{\lambda^3}{g}\right)^3.
\end{equation}
For $n=5$: 
\begin{eqnarray}
B_5&=&\left(\frac{5400\sqrt{30}+7925\sqrt{15}-25200\sqrt{5}-6912\sqrt{3}}{43200\sqrt{15}}\right)\left(\frac{\lambda^3}{g^2}\right)^4\nonumber\\
&=&\left(\frac{317}{1728}+\frac{\sqrt{2}}{8}-\frac{7\sqrt{3}}{36}-\frac{4\sqrt{5}}{125}\right)\left(\frac{\lambda^3}{g}\right)^4
\end{eqnarray}
For $n=6$: 
\begin{equation}
B_6=\left(\frac{23}{128}+\frac{2081\sqrt{2}}{6912}-\frac{\sqrt{3}}{72}-\frac{91\sqrt{6}}{432}-\frac{\sqrt{10}}{20}\right)\left(\frac{\lambda^3}{g}\right)^5.
\end{equation}
For $n=7$: 
\begin{equation}
B_7=\left(\frac{5957}{6912}+\frac{9\sqrt{2}}{64}-\frac{1721\sqrt{3}}{3888}-\frac{4\sqrt{5}}{25}-\frac{\sqrt{6}}{12}-\frac{6\sqrt{7}}{343}\right)\left(\frac{\lambda^3}{g}\right)^6.
\end{equation}

\begin{table}[!ht]
\begin{center}
\begin{tabular}{|c|c|c|}\hline
 Order $n$ & Virial coefficient $B_n\left(\frac{g}{\lambda^3}\right)^{n-1}$ & Numerical value \\\hline
       &   & \\ 
 $1$ & 1 & 1 \\
       &   & \\  
$2$ & $\cfrac{1}{2^{5/2}}$ & 0.176777 \\
       &   & \\
$3$ & $\cfrac{1}{8}-\frac{2}{9\sqrt{3}}$ & -0.00330006 \\
       &   & \\
$4$ & $\cfrac{3\sqrt{6}+5\sqrt{3}-16}{32\sqrt{6}}$ & 0.000111289 \\
       &   & \\
$5$ & $\cfrac{317}{1728}+\cfrac{\sqrt{2}}{8}-\cfrac{7\sqrt{3}}{36}-\cfrac{4\sqrt{5}}{125}$ & -0.0481161 \\
       &   & \\
$6$ & $\cfrac{23}{128}+\cfrac{2081\sqrt{2}}{6912}-\cfrac{\sqrt{3}}{72}-\cfrac{91\sqrt{6}}{432}-\cfrac{\sqrt{10}}{20}$ & -0.092685 \\
       &   & \\
$7$ & $\cfrac{5957}{6912}+\cfrac{9\sqrt{2}}{64}-\cfrac{1721\sqrt{3}}{3888}-\cfrac{4\sqrt{5}}{25}-\cfrac{\sqrt{6}}{12}-\cfrac{6\sqrt{7}}{343}$ & -0.31415 \\
       &   & \\\hline
\end{tabular}
\end{center}
\end{table}

\section{Analytical formula}

Setting, keeping Kilpatrick's notation
\begin{equation}\label{pj}
p_j=\frac{g}{\lambda^3}\frac{(-1)^{j+1}}{j^{3/2}},
\end{equation}
we have the relation
\begin{equation}
\sum_{j=1}^{\infty}\frac{p_j}{j}z^j=\sum_{k=1}^{\infty}B_k\left(\sum_{j=1}^{\infty}p_jz^j\right)^k
\end{equation}
and therefore
\begin{equation}
p_n=\frac{n}{2\pi i}\oint\frac{1}{z^{n+1}}\sum_{k=1}^{\infty}B_k\left[\sum_{j=1}^{\infty}p_jz^j\right]^kdz
\end{equation}
yielding
\begin{equation}\label{pn}
p_n=n\sum_{i=1}^ni!B_i\sum_{\{r_s\}}\prod_{s=1}^n\frac{p_s^{r_s}}{r_s!}
\end{equation}
with
\begin{equation}
\sum_{s=1}^nr_s=i
\end{equation}
and
\begin{equation}
\sum_{s=1}^nsr_s=n.
\end{equation}
For instance, in the case $n=3$, one has
\begin{equation}
\left\{
\begin{array}{l}
p_1=p_1B_1\\
\frac{1}{2}p_2=p_2B_1+p_1^2B_2\\
\frac{1}{3}p_3=p_3B_1+2p_2p_1B_2+p_1^3B_3.
\end{array}\right.
\end{equation}
In order to express the $B_k$ coefficients in terms of the $p_j$, let us write
\begin{equation}
\sum_{j=1}^{\infty}\frac{p_j}{j}z^j=\sum_{k=1}^{\infty}B_k\rho^k
\end{equation}
Integrating over $\rho$ after multiplication by $\rho^{-n+1}$ yields
\begin{equation}
B_n=\frac{1}{2\pi i}\oint\sum_{j=1}^{\infty}\frac{p_j}{j}z^j\times\frac{1}{\rho^{n-1}}d\rho.
\end{equation}
Using
\begin{equation}
\rho=\sum_{j=1}^{\infty}p_jz^j,
\end{equation}
one gets
\begin{equation}
B_n=\frac{1}{2\pi i}\oint\left(\sum_{j=1}^{\infty}\frac{p_j}{j}z^j\right)\left(\sum_{k=1}^{\infty}p_kz^k\right)^{-n-1}\left(\sum_{l=1}^{\infty}lp_lz^{l-1}\right)dz.
\end{equation}
Thus, $B_n$ is the coefficient of $z^n$ in the expansion of
\begin{equation}
B_n=\frac{1}{2\pi i}\oint\left(p_1z\right)^{-n-1}\left(1+\sum_{k=2}^{\infty}\frac{p_k}{p_1}z^{k-1}\right)^{-n-1}\left(\sum_{j=1}^{\infty}\frac{p_j}{j}z^j\right)\left(\sum_{l=1}^{\infty}lp_lz^{l-1}\right)dz.
\end{equation}
and $B_n$ is the coefficient of $z^n$ in the expansion of
\begin{equation}
p_1^{-n-1}\left(1+\sum_{k=2}^{\infty}\frac{p_k}{p_1}z^{k-1}\right)^{-n-1}\left(\sum_{j=1}^{\infty}\frac{p_j}{j}z^j\right)\left(\sum_{l=1}^{\infty}lp_lz^{l-1}\right)
\end{equation}
Expanding the different terms, one gets
\begin{equation}
B_n=\sum_{i=0}\sum_{j=1}\sum_{k=1}\frac{(-1)^i(n+i)!}{n!p_1^{n+1+i}}\frac{kp_jp_k}{j}\sum_{\{r_s\}}\prod_{s=2}^n\frac{p_s^{r_s}}{r_s!}
\end{equation}
with
\begin{equation}
\sum_{s=2}^nr_s=i
\end{equation}
and
\begin{equation}
\sum_{s=2}^nsr_s=n+i+1-j-k.
\end{equation}
As shown by Kilpatrick \cite{Kilpatrick1953,Putnam1953}, one can set $k'_s=r_s$ for $s\ge 2$, $k''_s=\delta_{sj}$, $k'''_s=\delta_{sk}$ and $k_s=k'_s+k''_s+k'''_s$. One has subsequently
\begin{equation}
\sum_{s=2}^nk_s=i+2-k''_1-k'''_1
\end{equation}
and
\begin{equation}
\sum_{s=2}^nsk_s=n+i+1-k''_1-k'''_1.
\end{equation}
Concerning the factor in $p_1$, the largest possible value of $i$ is $n-1$, since no larger integer can divide $n-1+i$ into $i$ parts, each one having size larger or equal than 2. For that reason, Kilpatrick suggested to write $p_1^{2n-2}$ in the denominator and the remaining term in the form $p_1^{k_1}$. This defines $k_1$, and therefore $k'_1$. One gets
\begin{equation}
-n-1-i+k''_1+k'''_1=k_1-(2n-2)
\end{equation}
i. e. $i=n-3-k'_1$. The two constraints on the summation thus read
\begin{empheq}[box=\fbox]{align*}
\sum_{s=1}k_s=n-1
\end{empheq}
and
\begin{empheq}[box=\fbox]{align*}
\sum_{s=1}sk_s=2n-2
\end{empheq}
and
\begin{empheq}[box=\fbox]{align*}
B_n=\sum_{\{k_s\}}\frac{(-1)^{n-1-k_1}(n-1)(2n-k_1-3)!}{n!p_1^{2n-2}}p_1^{k_1}\prod_{s=2}^n\frac{p_s^{k_s}}{k_s!}.
\end{empheq}
Kilpatrick pointed out at the end of its paper that $B_n$ is in fact the coefficient of $z^{2n}$ in the expansion of
\begin{equation}
\frac{1}{n}\sum_{j=1}^{\infty}\left(p_jz^j\right)^{-n+1}.
\end{equation}
Such a property can probably be useful in order to find the expression of $B_n$, using Fa\`a di Bruno and multinomial coefficients.
Replacing $p_j$ by its value (\ref{pj}) in our specific case, one gets
\begin{empheq}[box=\fbox]{align*}
B_n=&\left(\frac{\lambda^3}{g}\right)^{n-1}\frac{n-1}{n!}\sum_{\{k_s\}}(2n-k_1-3)!\frac{(-1)^{k_1}}{\prod_{s=2}^nk_s!s^{3k_s/2}}.
\end{empheq}
For instance, the first five coefficients are
\begin{equation}
\left\{
\begin{array}{l}
B_1=1\\
B_2=-\frac{p_2}{p_1^2}\\
B_3=\frac{1}{p_1^4}\left(-\frac{2}{3}p_3p_1+p_2^2\right)\\
B_4=\frac{1}{p_1^6}\left(-\frac{3}{4}p_4p_1^2+3p_3p_2p_1-\frac{5}{2}p_1^3\right)\\
B_5=\frac{1}{p_1^8}\left(-\frac{4}{5}p_5p_1^3+4p_4p_2p_1^2+2p_3^2p_1^2-12p_3p_2^2p_1+7p_2^4\right)
\end{array}\right..
\end{equation}
Note that
\begin{equation}
\prod_{s=2}^nk_s!
\end{equation}
is $G(n+2)$ where $G(z)$ represents the Barnes $G$ function
\begin{equation}
G(n)=\prod_{k=1}^n\Gamma(k).
\end{equation}

\section{Sum rules}

Equation (\ref{pn}) becomes
%
%
\begin{equation}\label{sr}
\sum_{i=1}^n(-1)^{i-1}i!\left(\frac{g}{\lambda^3}\right)^{i-1}B_i\sum_{\{r_s\}}\frac{1}{\prod_{s=1}^n\left[r_s!\times s^{3r_s/2}\right]}=\frac{1}{n^{5/2}}
\end{equation}
with
\begin{equation}
\sum_{s=1}^nr_s=i\;\;\;\;\mathrm{and}\;\;\;\;\sum_{s=1}^nsr_s=n.
\end{equation}
Equation (\ref{sr}) constitutes a sum rule which can be useful to check numerical calculations of the $B_n$ coefficients. It can also be used to express $B_n$ in terms of the $B_i$, $i\leq n-1$ as:
\begin{eqnarray}
B_n&=&\frac{(-1)^{n-1}}{n!}\left(\frac{\lambda^3}{g}\right)^{n-1}\times\left(\sum_{\{q_s^{(n)}\}}\frac{1}{\prod_{s=1}^n\left[q_s^{(n)}\times s^{3q_s^{(n)}/2}\right]}\right)^{-1}\nonumber\\
& &\times\left[\frac{1}{n^{5/2}}-\sum_{i=1}^{n-1}(-1)^{i-1}i!\left(\frac{g}{\lambda^3}\right)^{i-1}B_i\sum_{\{q_s^{(i)}\}}\frac{1}{\prod_{s=1}^n\left[q_s^{(i)}\times s^{3q_s^{(i)}/2}\right]}\right],
\end{eqnarray}
where
\begin{equation}
\sum_{s=1}^nq_s^{(i)}=i
\end{equation}
and
\begin{equation}
\sum_{s=1}^nsq_s^{(i)}=n.
\end{equation}

\section{Conclusion}

In this document, we proposed a discussion about the coefficients of the Virial expansion. We followed the general derivation of Kilpatrick to obtain analytical expressions for the Fermi ideal gas. It is worth mentioning that Wilson and Rogers presented relations in the cluster expansion theory of non-ideal gases using the formalism of umbral calculus \cite{Wilson1986}.

\end{document}